\documentclass[12pt]{article}

\usepackage{graphicx,setspace}
\usepackage{multicol,multirow}
\usepackage{amsmath,amssymb,amsfonts}
\usepackage{mathrsfs}
\usepackage{amsthm}
\usepackage{rotating}
\usepackage{appendix}
\usepackage{ifpdf}
\usepackage[T1]{fontenc}
\usepackage{newtxtext}
\usepackage{newtxmath}
\usepackage{textcomp}
\usepackage{xcolor}
\usepackage{lipsum}
\usepackage[colorlinks,allcolors=blue]{hyperref}

\theoremstyle{definition}

\numberwithin{equation}{section}

\usepackage{ulem,comment}
\usepackage{booktabs}
\usepackage{caption}
\usepackage{subcaption}
\usepackage{fullpage}
\usepackage{natbib}
\usepackage{bm}
\usepackage{amsmath}
\usepackage{graphicx}
\usepackage{xcolor}

\usepackage{longtable}

\newcommand{\ind}{\stackrel{\text{ind}}{\sim}}

\newcommand{\E}{\text{E}}

\newcommand{\bx}{\boldsymbol{x}}

\newcommand{\bbeta}{\boldsymbol{\beta}}

\newcommand{\bg}{\boldsymbol{\beta_\gamma}}

\newcommand{\ga}{\boldsymbol{\gamma}}
\newcommand{\G}{\boldsymbol{\Gamma}}
\newcommand{\bicpi}{\hat{\pi}}
 
\graphicspath{{plots/}}
\providecommand{\keywords}[1]{\small\textbf{Keywords:} #1}

\begin{document}

\title{Online Bayesian Model Averaging with Joint Uncertainty Quantification for Models and Regression Coefficients in Binary Regression}

\author{Joyee Ghosh\footnote{Associate Professor, Department of Statistics and Actuarial Science, The University of Iowa, Iowa City, USA}
 \ and Aixin Tan\footnote{Associate Professor, Department of Statistics and Actuarial Science, The University of Iowa, Iowa City, USA}
 }

\date{}

\maketitle

\begin{abstract}
Streaming binary response data arise in many applications, including virtual learning platforms where student responses are collected sequentially and estimated probabilities of correct responses may inform future question assignment. Logistic regression provides an interpretable framework for such analysis, but the data may support multiple plausible predictor subsets. Bayesian model averaging (BMA) accounts for such model uncertainty by averaging inference across competing models. Since repeatedly applying BMA to accumulating data can be computationally burdensome, we develop an online implementation based on renewable estimation. At each update, retained summaries and new observations are used to approximate the joint posterior of models and model-specific coefficients without revisiting historical data, enabling point and interval estimates that incorporate model uncertainty. Simulations show that online BMA closely approximates its offline counterpart in coefficient estimation, model and variable importance, prediction, and interval estimation of predictive probabilities, while substantially reducing computational cost. In a virtual learning application, we find substantial uncertainty across competing predictor subsets and large variability in credible interval widths across questions. Our method distinguishes questions with precise estimates from those with substantial uncertainty, providing information beyond point estimates for adaptive question assignment.
\\
\\

\end{abstract}

\doublespacing

\section{Introduction}
The increasing prevalence of streaming data has created a growing need for statistical methods that update inference efficiently as new observations arrive. Virtual learning platforms provide one representative example, where student responses are recorded continuously over extended periods. Such data enable ongoing estimation of quantities such as the probability that a student correctly answers a question, which may help inform future question assignment. Besides accurate point estimates, quantifying uncertainty is equally important, since two questions with similar estimated success probabilities may differ substantially in the reliability of those estimates. In this paper, we develop a BMA method for streaming binary response data that efficiently updates inference while providing uncertainty quantification, without repeated analysis of the accumulated dataset. The proposed framework characterizes uncertainty in model selection, predictor importance, and prediction through posterior model probabilities, marginal posterior inclusion probabilities for individual predictors, and credible intervals for quantities of interest, such as the probability of a successful response.

We begin with the classical offline setting, where all data are available. Logistic regression provides a natural and interpretable framework for binary responses. In educational applications, predictors may include student and problem characteristics together with features summarizing prior performance and accumulated learning opportunities \citep{pav:egli:harr:2021}. With multiple potentially informative predictors, there may be considerable uncertainty about which predictor subset best describes the response. 

Standard variable selection approaches
typically select a single model and proceed as though it were known,
potentially discarding information from other well supported models.
Instead of choosing a single model, BMA averages inference over competing models according to their posterior probabilities  \citep{Hoet:Madi:Raft:Voli:1999}, accounting for uncertainty about which predictors should be included.

\begin{figure}
    \centering 
 \includegraphics
 [height=3.3in,width=6.6in]{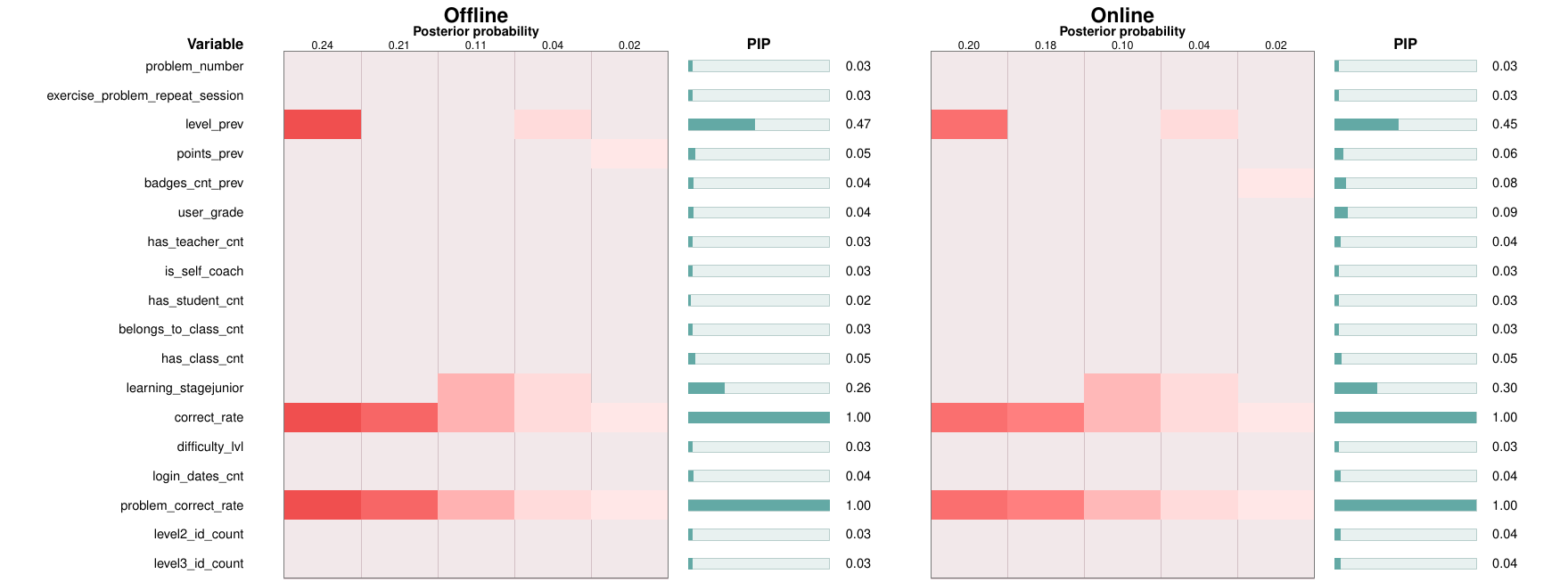}   
\caption{
Display of offline and online BMA 
results for variable and model importance for Junyi data.  Online method results closely approximate that of offline. (The intensity of the red color reflects the magnitude of the model posterior probability). }
    \label{fig:junyi-topmodels}
\end{figure}

The Junyi Academy data analyzed in Section~\ref{sec:application} illustrates such model uncertainty. As shown in Figure~\ref{fig:junyi-topmodels}, the three highest probability models under online BMA receive posterior probabilities of only $0.20$, $0.18$, and $0.10$, respectively, with the remaining probability distributed across
many additional models. Thus, an analysis based only on the 
highest probability model alone would condition inference on a predictor
configuration receiving only $20\%$ of the posterior support and discard the
remaining $80\%$. BMA instead retains information from all competing models
according to their posterior support. It also provides marginal posterior
inclusion probabilities (PIPs), summarizing the importance of individual
predictors.
  
  The price to pay for the rich inference from BMA is the computing time: the offline version was unbearably costly and the cost grows as data accumulates. We therefore develop an online implementation designed to reduce this burden while preserving, as closely as possible, the inferential accuracy of the corresponding offline analysis.

Several computational strategies can facilitate inference as data accumulate, including subsampling, divide-and-conquer methods, and sequential updating \citep{wang:2016}. The first two reduce computing burden but typically require access to all historical data. In contrast, the online updating strategy processes data sequentially in batches and retains only limited summary statistics, making it particularly attractive for long-running data streams. Frequentist online methods have been developed for quantile regression and generalized linear models, offering scalable point estimation and asymptotic inference under a fixed model \citep{schifano:2016,wang:2022,fan:lin:2025}.

Bayesian approaches to online inference remain comparatively underdeveloped, especially for variable selection problems. Building on renewable estimation ideas, \citet{ghosh:tan:luo:2025} proposed an online Bayesian logistic regression framework that incorporates model uncertainty through BMA using marginal likelihood approximations based on the Bayesian information criterion (BIC) \citep{schwarz:1978}. While this approach enables efficient online model selection and averaging without storing the full data, it focuses primarily on posterior model probabilities and does not provide the model specific posterior distributions of the regression coefficients. 

In this paper, we extend this framework to approximate the joint posterior distribution of the model indicator and model specific regression coefficients for streaming binary response data. A key feature of our proposed technique is that the same renewable summaries used to update posterior model probabilities also provide the quantities needed for a Gaussian approximation to the model specific posterior distribution of the logistic regression coefficients. Our online BMA method therefore requires neither storage nor reanalysis of historical observations and permits independent Monte Carlo sampling from the approximate joint posterior. The method provides, at each online update, posterior model probabilities, PIPs, predictor effect sizes, and predicted success probabilities with associated credible intervals.

\section{Online BMA for Logistic Regression}
\label{sec:method}

\subsection{BMA  for logistic regression}\label{subsec:bma}

Let $y_i\in\{0,1\}$ denote a binary response and $\bx_i=(x_{i1},\ldots,x_{ip})^T$ the corresponding vector of candidate predictors, for $i=1,\ldots,n$. We introduce a model indicator $\ga=(\gamma_1,\ldots,\gamma_p)^T$, where $\gamma_j=1$ if the $j$th predictor is included and $\gamma_j=0$ otherwise. The collection of candidate models is denoted by $\G$. For a given model $\ga\in\G$, let $\bx_{i,\ga}$ contain the predictors selected by $\ga$, and let $\bg$ be the associated $q_{\ga}$-dimensional regression coefficient vector, where $q_{\ga}=\sum_{j=1}^p\gamma_j$.

Under model $\ga$, we assume
\begin{equation}
\label{eq:logistic-model}
y_i\mid \bx_i,\bg,\ga \ind \mathrm{Bernoulli}(\mu_{i,\ga}),
\qquad
\operatorname{logit}(\mu_{i,\ga})=\bx_{i,\ga}^{T}\bg.
\end{equation}
For the dataset $D=\{(\bx_i,y_i):i=1,\ldots,n\}$, the log-likelihood under model $\ga$ is
\begin{equation}
\label{eq:loglik-gamma}
l_{\ga}(\bg;D)
=\sum_{i=1}^n\left\{y_i\bx_{i,\ga}^{T}\bg-
\log\left(1+\exp(\bx_{i,\ga}^{T}\bg)\right)\right\}.
\end{equation}
The Bayesian specification then assigns priors
\[
\bg\mid\ga\sim p(\bg\mid\ga),
\qquad
\ga\sim p(\ga).
\]
The prior on $\bg$ can incorporate historical information or expert knowledge when available. It can also stabilize coefficient estimation under severe
multicollinearity, when the frequentist maximum likelihood estimate (MLE) is highly variable, and provide inference under separation, when MLEs do not exist. The prior on $\ga$ can similarly encode prior knowledge or preferences regarding predictor inclusion, thereby
guiding model selection and restricting the candidate
model space for computational or practical considerations.

The inferential target is the joint posterior distribution
\begin{equation}
\label{eq:joint-posterior}
\pi(\ga,\bg\mid D)
\propto
\exp\{l_{\ga}(\bg;D)\}
 p(\bg\mid\ga)p(\ga),
\end{equation}
which allows simultaneous inference on model uncertainty and regression parameters. It admits the factorization
\begin{equation}
\label{eq:joint-factorization}
\pi(\ga,\bg\mid D)
=
\pi(\bg\mid\ga,D)\pi(\ga\mid D),
\end{equation}
where
\begin{equation}
\label{eq:conditional-beta}
\pi(\bg\mid\ga,D)
=
\frac{\exp\{l_{\ga}(\bg;D)\}p(\bg\mid\ga)}
{m(D\mid\ga)}\,,
\end{equation}
and
\begin{equation}
\label{eq:posterior-model}
\pi(\ga\mid D)
=
\frac{m(D\mid\ga)p(\ga)}
{\sum_{\ga'\in\G}m(D\mid\ga')p(\ga')}\,.
\end{equation}
Here,
\begin{equation}
\label{eq:marginal-likelihood}
m(D\mid\ga)
=
\int \exp\{l_{\ga}(\bg;D)\}p(\bg\mid\ga)\,d\bg
\end{equation}
is the marginal likelihood under model $\ga$.

Bayesian posterior inference generally requires evaluating integrals with respect to the joint posterior distribution in \eqref{eq:joint-factorization}. For logistic regression, the model-specific posterior is available up to a normalizing constant, but evaluating posterior expectations and the marginal likelihood in \eqref{eq:marginal-likelihood} involves intractable integration. Standard approaches include numerical or asymptotic approximations
and MCMC methods; see, for example, \citet{Raft:1996:bka} and
\citet{polson:scott:windle:2013}. When these calculations are needed for many candidate models and repeatedly as data accumulate,
computationally simpler approximations are preferred. One widely used large-sample approximation to the posterior model probability is based on the Bayesian information criterion (BIC) \citep{schwarz:1978,Raft:1996:bka}, and is given by
\begin{equation}
\label{eq:pgamma-bic}
\bicpi(\ga|D)=
\frac{\exp\{-\frac{1}{2}\textrm{BIC}(\ga; D)\}p(\ga)}
{\sum_{\ga'\in\G}
\exp\{-\frac{1}{2}\textrm{BIC}(\ga', D)\}p(\ga')}\,,
\end{equation}
where
\begin{equation}
\label{eq:BIC}
\textrm{BIC}(\ga;D)
=
-2l(\hat{\bbeta}_{\ga};D)
+
q_{\ga}\log(n),
\end{equation}
can be computed from the MLE $\hat{\bbeta}_{\ga}$, its corresponding log-likelihood $l(\hat{\bbeta}_{\ga};D)$, and the number of
predictors $q_{\ga}$ under model $\ga$.  
As shown below, this BIC approach is suitable for extension to online updating.  We refer to its classical
implementation to accumulated data as Offline BMA, which will serve as a benchmark for
the online method developed below.

\subsection{Online BMA with approximation to the joint posterior}
\label{sec:online-joint}

In the streaming setup, we seek efficient ways to do BMA at each time point via the joint distribution based on \eqref{eq:conditional-beta} and \eqref{eq:pgamma-bic}. Suppose the full data $D$ arrive sequentially in batches
$D_1,\ldots, D_B$, and let
$D_b^*=\{D_1,\ldots,D_b\} $
denote the accumulated data by batch $b$, with cumulative sample size
$N_b$, for $b=1,\ldots, B$.

At time $b$,  \eqref{eq:pgamma-bic} depends on 
\begin{equation}\label{eq:BICb}
\mathrm{BIC}(\ga;D_b^*)
=
-2\,l(\hat{\bbeta}_{\ga};D_b^*)
+
q_{\ga}\log N_b\,,
\end{equation}
for $\ga\in\G$, the direct evaluation of which would require retaining and repeatedly processing all observations in $D_b^*$. 
For efficient online estimation of $\hat{\bbeta}_{\ga}$,
\citet{Luo_2020} proposed a renewable estimation algorithm that
updates the MLE recursively by using only the current batch, $D_b$, together with stored
summary statistics from previous batches, $D_1$ to $D_{b-1}$. We denote the resulting
renewable estimator by $\tilde{\bbeta}_{\ga}$.
For BMA, however, the BIC in \eqref{eq:BICb}
requires the accumulated log-likelihood
$l(\tilde{\bbeta}_{\ga};D_b^*)$.
Because 
$D_1,\ldots,D_{b-1}$ have been discarded, this quantity cannot be
computed directly, motivating two recursive log-likelihood
approximations developed by 
\citet{ghosh:tan:luo:2025}.  
The first approximation, Online~1, is
\[
l_{\mathrm{On1}}
(\tilde{\bbeta}_{\ga;b};D_b^*)
=
\sum_{k=1}^{b}
l(\tilde{\bbeta}_{\ga;k};D_k)=l_{\mathrm{On1}}
(\tilde{\bbeta}_{\ga;b-1};D_{b-1}^*)
+
l(\tilde{\bbeta}_{\ga;b};D_b)\,,
\]
where 
$\left(\tilde{\bbeta}_{\ga;k},
l(\tilde{\bbeta}_{\ga;k};D_k)\right)$, the renewable estimator and its likelihood, serve as summaries to be stored after each batch up to the current time, in place of the full history.

Online~2 improves this approximation by incorporating a quadratic
correction based on the observed Hessian. Specifically,
\[
\begin{aligned}
l_{\mathrm{On2}}
(\tilde{\bbeta}_{\ga;b};D_b^*)
&=
l_{\mathrm{On2}}
(\tilde{\bbeta}_{\ga;b-1};D_{b-1}^*)
+
l(\tilde{\bbeta}_{\ga;b};D_b)
\\
&\quad
-\frac12
(\tilde{\bbeta}_{\ga;b}
-
\tilde{\bbeta}_{\ga;b-1})^\top
\widetilde{J}_{\ga;b-1}
(\tilde{\bbeta}_{\ga;b}
-
\tilde{\bbeta}_{\ga;b-1})\,.
\end{aligned}
\]
Here, $\widetilde{J}_{\ga;k-1}$ is the accumulated observed information matrix that is readily available from the renewable estimation algorithm. After batch $k$, the retained summaries are \begin{equation}\label{eq:summary}\left(\tilde{\bbeta}_{\ga;k},\,
l(\tilde{\bbeta}_{\ga;k};D_k),\,\widetilde{J}_{\ga;k}\right)\,.
\end{equation}
The Online~2 approximation to the accumulated log-likelihood is then substituted into the BIC expression in \eqref{eq:BICb} after the arrival
of batch $b$, yielding the approximate posterior model probabilities
\begin{equation}\label{eq:online-model-prob}\widetilde{\pi}(\ga|D^*_b)\,.\end{equation}
We adopt Online~2 throughout this paper due to its higher accuracy than Online~1.

The BIC approximation is evaluated for each $\ga\in\Gamma$, which is feasible when the candidate model space $\G$ can be enumerated. For substantially
larger model spaces, stochastic model search can be used to identify a
manageable pool of promising models
\citep{Raft:Madi:Hoet:1997,ghosh:tan:luo:2025}, with the required renewable
summaries maintained for models in this pool as data arrive. 

To extend the approximate posterior model probabilities of
\citet{ghosh:tan:luo:2025} to the joint posterior distribution,
BMA additionally requires the model-specific
conditional posterior distributions
$\pi(\bbeta_{\ga}\mid\ga,D_b^*)$ from \eqref{eq:conditional-beta} for $D=D_b^*$ at each $b$.
Like the accumulated log-likelihood, these conditional posteriors cannot
be evaluated directly because observations from previous batches are no
longer available.
The key observation underlying the present work is that the renewable
mode $\widetilde{\bbeta}_{\ga,b}$ and observed information matrix
$\widetilde{J}_{\ga,b}$ maintained by the Online~2 algorithm contain
precisely the information needed to construct a Gaussian approximation
to the conditional posterior.
Consequently, no additional passes through the historical data are
required, and we approximate
\begin{equation}
\label{eq:online-cond-post}
\widetilde{\pi}(\bbeta_{\ga}\mid\ga,D_b^*)
\approx
\mathcal{N}_{q_{\ga}}
\left(
\widetilde{\bbeta}_{\ga,b},
\widetilde{J}_{\ga,b}^{-1}
\right).
\end{equation}

Combining \eqref{eq:online-model-prob} and
\eqref{eq:online-cond-post} therefore gives the approximate joint posterior
\begin{equation}
\label{eq:online-joint-post}
\widetilde{\pi}(\ga,\bbeta_{\ga} | D_b^*)=
\widetilde{\pi}(\bbeta_{\ga} | \ga,D_b^*)
\widetilde{\pi}(\ga | D_b^*).
\end{equation}

Our proposed Gaussian approximation is justified in the streaming data setting due to the availability of large samples and is computationally attractive. 
In particular, we use independent Monte Carlo sampling for computation. At batch $b$, each
joint posterior draw is obtained by first sampling $\ga$ from
\eqref{eq:online-model-prob} and then sampling $\bg$ conditionally on $\ga$ from
\eqref{eq:online-cond-post}. MCMC methods, such
as Polya--Gamma data augmentation \citep{polson:scott:windle:2013} and
pSUN-based methods \citep{onor:lise:2025}, target the logistic regression posterior  directly but are more computationally demanding,  may encounter mixing issues, and is non-trivial to extend to an online implementation. In contrast, our approach reuses the summaries already maintained by Online~2 to approximate the joint
posterior without revisiting historical data, while
avoiding the need to assess MCMC convergence and mixing.

\subsection{Statistical inference}
\label{subsec:statinf}
 We now present inferential tools enabled by our online BMA framework. In particular, the joint posterior distribution of the model and model-specific coefficients enables inference beyond that available from previous online methods based solely on posterior model probabilities. Below, let $(\bg^{[t]},\ga^{[t]}, t=1,\ldots,T)$ denote the i.i.d.\ Monte Carlo draws from the approximate joint posterior $\widetilde{\pi}$, with the batch index suppressed for notational simplicity.\\\vspace{.5cm}

\noindent \textbf{Predictive probability that accounts for model uncertainty.} For a new covariate vector $\bx_{\mathrm{new}}$, we may be interested in both a point estimate and a credible interval for its success probability. We first compute, for each Monte Carlo draw,
\begin{equation}
    p^{[t]}=\frac{exp{\left(\bx_{\mathrm{new},\ga^{[t]}}^{T}\,\bg^{[t]}\right)}}{1+exp{\left(\bx_{\mathrm{new},\ga^{[t]}}^{T}\,\bg^{[t]}\right)}}\,.
    \label{eq:predictiveprob}
\end{equation}

Then the sample mean and quantiles of $\{p^{[t]}:t=1,\ldots,T\}$ are Monte Carlo estimates of the posterior mean and posterior quantiles of the success probability, respectively. The former is used as the point estimate of the success probability, and the latter are used to form the posterior credible interval. 
These summaries incorporate both uncertainty in the
model-specific regression coefficients and uncertainty across competing
models.\\

\vspace{.5cm}
\noindent \textbf{Model importance and variable importance.} 
Posterior model probabilities directly quantify support for all the candidate predictor configurations. The  marginal posterior inclusion probability
(PIP) of predictor $j$ is 
\begin{equation}
\label{eq:pip}
p(\gamma_j=1\mid D)
\approx
\sum_{\ga\in\G:\gamma_j=1}\tilde{\pi}(\ga\mid D).
\end{equation}
Alternatively, one can use directly their respective Monte Carlo estimates, $\frac{1}{T}\sum_{t=1}^T\ga^{[t]}$, and $\frac{1}{T}\sum_{t=1}^T\ga_j^{[t]}$. \\

\vspace{.5cm}
\noindent \textbf{Variable effect size.} To summarize regression effects across models, define the $p$-dimensional coefficient vector $\bbeta^{(\ga)}$ by placing the elements of $\bg$ in
their corresponding positions and setting the coefficients of excluded predictors to zero. The model averaged posterior mean is then approximated by
\begin{equation}
\label{eq:bma-beta-mean}
\E(\bbeta\mid D) \approx \sum_{\ga\in\G}
\widetilde{\pi}(\ga\mid D)\,
\E\{\bbeta^{(\ga)}\mid\ga,D\},
\end{equation}
with Monte Carlo estimate
$\frac{1}{T}\sum_{t=1}^T{\bbeta^{(\ga)}}^{[t]}$.

For predictor $j$, its marginal posterior mean, $\E(\beta_j\mid D)$, provides one way to summarize its model averaged effect, where $\exp\{\E(\beta_j\mid D)\}$ may be interpreted as the odds ratio associated with a one unit increase in $x_j$, while fixing the other predictors. The posterior of $\beta_j$ is indeed a mixture of a point mass at zero, corresponding to models that exclude the predictor, and a continuous component from models that
include it, making conventional credible intervals hard to interpret. One may instead condition on one of the leading models containing predictor $j$ and use the conditional posterior mean or median to summarize its effect, while, for example, use the 5\% and 95\% quantiles of the corresponding Monte Carlo draws to estimate a 90\% credible interval. More generally, the joint conditional posterior $\pi(\bbeta_{\ga}\mid\ga,D)$ also provides simultaneous inference for the coefficients included in model $\ga$ and can be used to construct a
joint credible region for $\bbeta_{\ga}$.\\

\section{Simulation Studies}\label{sec:simulation} 
In this section, we conduct simulation studies to compare the performance of 
our proposed online BMA method to its offline BMA counterpart. Our main goal is to demonstrate that the online BMA method maintains accuracy comparable to the corresponding offline method, while providing a substantial gain in speed. For all simulation settings, we consider $p=15$ predictor variables in addition to the intercept. We consider $B=101$ batches, with the first batch size $n_1=100$ and the remaining batch sizes $n_b=19$, for $b=2,3,\dots,B$, with aggregate sample size at the end of $B$ batches as $N_{B}=\sum_{b=1}^{B}n_b=2000$. For each of the $N_{B}$ observations, we independently generate the $p \times 1$ vector of predictors from a multivariate normal distribution with mean $\bm{0}$ and covariance matrix $\bm{\Sigma}$. 
Let $\bm{\rho}$ denote the corresponding correlation matrix. We consider 4 different settings by varying the correlation structure of the predictors, the sparsity level, and the magnitude of the regression coefficients. For the settings with  correlated predictors,  we consider a block diagonal correlation matrix to resemble the few blocks of highly and moderately correlated predictors in the Junyi data, given in Figure \ref{fig:junyi-corrX}. We consider $25$ replicates for each of the $4$ simulation scenarios. In all settings, we take $\bm{\Gamma}$ as the collection of top $5000$ models, determined by an offline enumeration of all $2^{p}=2^{15}=32768$ models in the first batch. We set our independent Monte Carlo sample size at 1000.

\begin{enumerate}
    \item Correlated sparse example: We consider a block diagonal correlation matrix $\bm{\rho}$ with ${\rho}_{1,2}=0.9, {\rho}_{3,4}=0.85, {\rho}_{5,6}=0.8, {\rho}_{7,8}={\rho}_{7,9}={\rho}_{8,9}=0.65$,
    and ${\rho}_{10,11}={\rho}_{10,12}=$ ${\rho}_{11,12}=0.25$. We take $\bm{\Sigma} = 3\bm{\rho}$. We take the intercept ${\beta}_0=1$, and 5 nonzero coefficients: $\beta_{3}=0.4$,
    $\beta_{4}=0.3$, $\beta_{10}=0.4$,
    $\beta_{12}=0.4$, and $\beta_{15}=0.5$. The rest of the $\beta_{j}$s are set to 0.
    
    \item Correlated nonsparse example: Here
    $\bm{\rho}$ is identical to the previous case and
     $\bm{\Sigma} = 1.5\bm{\rho}$. Again, the intercept ${\beta}_0$ is taken to be 1, and here we consider 8
     nonzero coefficients: $\beta_{5}=0.4$,
    $\beta_{6}=0.6$, $\beta_{7}=0.4$,
    $\beta_{8}=0.5$, $\beta_{9}=0.6$, $\beta_{13}=0.5$,
    $\beta_{14}=0.3$, and $\beta_{15}=1$. 
     
    \item Independent sparse example: Here we take $\bm{\rho}$ as the identity matrix and $\bm{\Sigma} = 3\bm{\rho}$. The regression coefficients are identical to the correlated sparse example.
    
    \item Independent nonsparse example: This scenario is identical to the correlated nonsparse case, except that $\bm{\rho}$ is the identity matrix and $\bm{\Sigma} = 1.5\bm{\rho}$.
\end{enumerate}

\begin{figure}
    \centering
\includegraphics
[width=\textwidth]{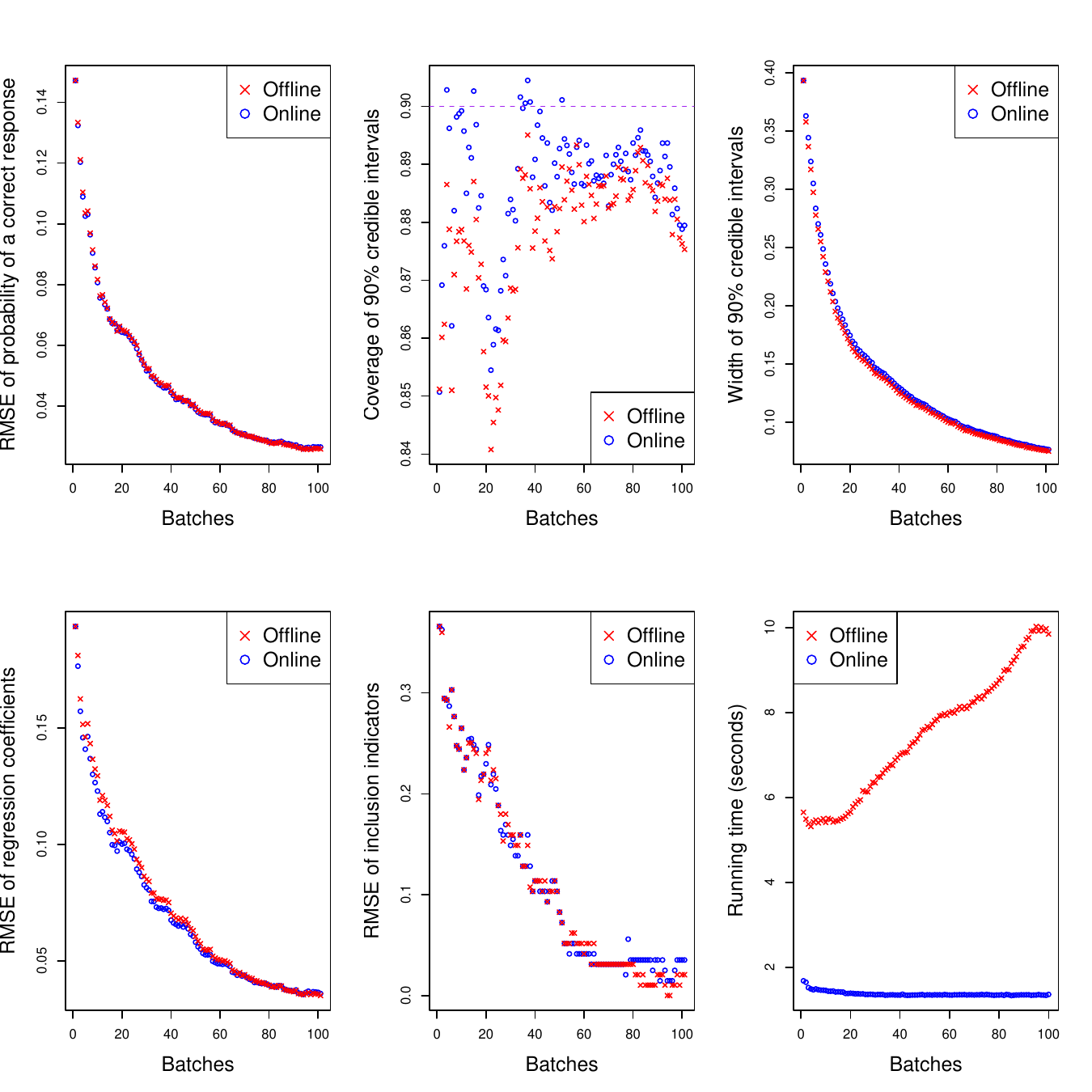}
    \caption{Simulation results for the correlated sparse example, averaged over 25 replicates.}
    \label{fig:corsparse}
\end{figure}

\begin{figure}
    \centering
\includegraphics
[width=\textwidth]{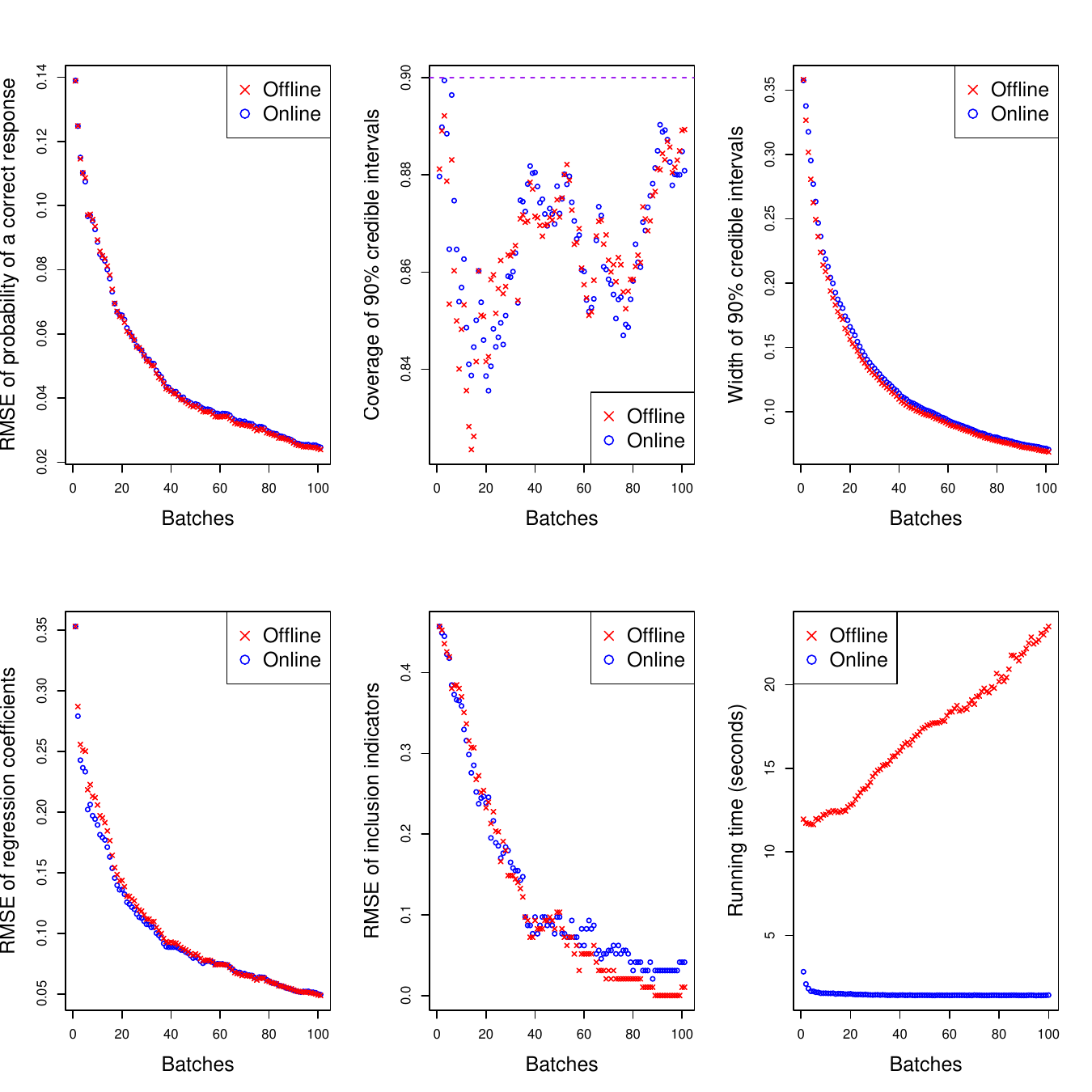}
    \caption{Simulation results for the correlated nonsparse example, averaged over 25 replicates.}
    \label{fig:cornonsparse}
\end{figure}

\begin{figure}
    \centering
\includegraphics
[width=\textwidth]{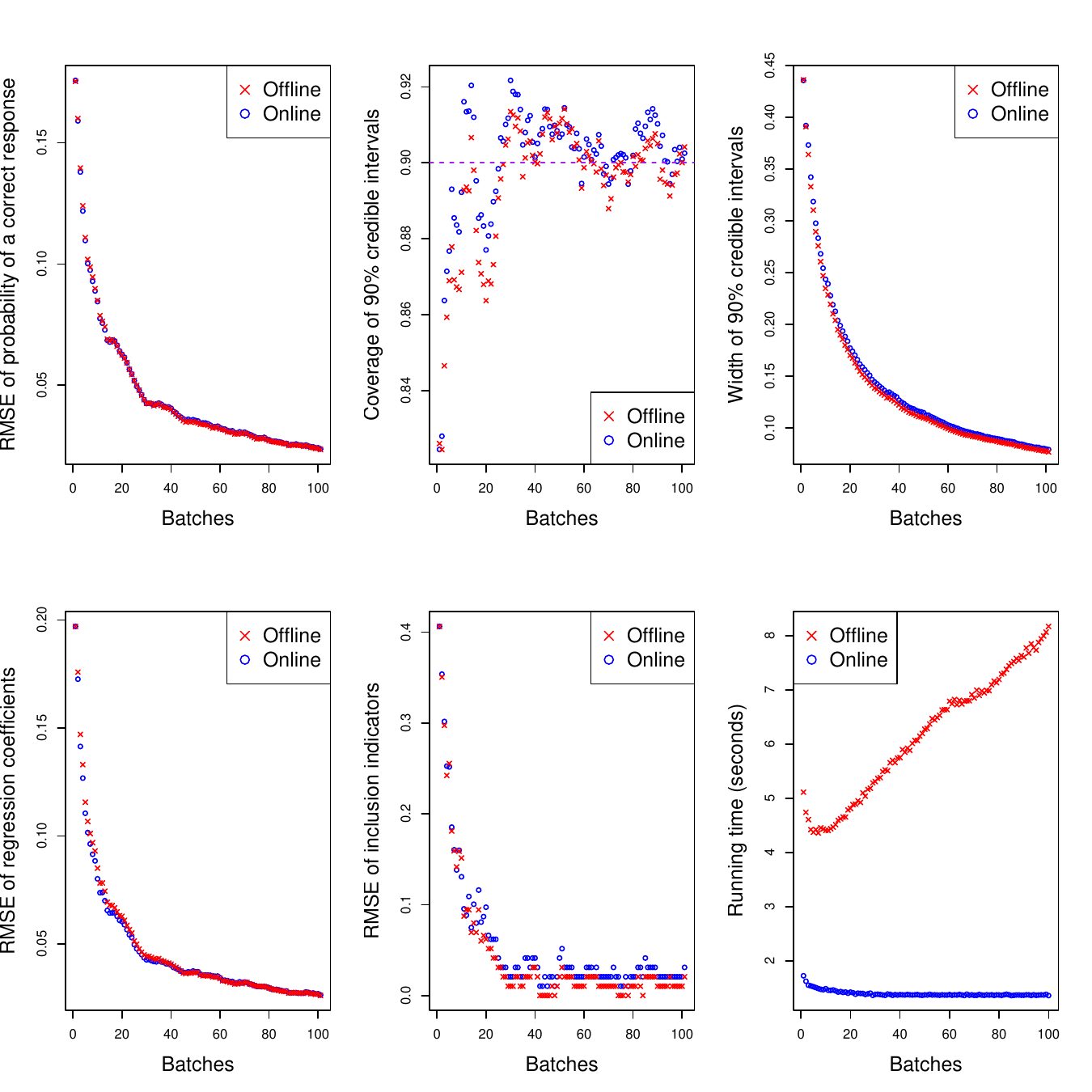}
    \caption{Simulation results for the independent sparse example, averaged over 25 replicates.}
    \label{fig:uncorsparse}
\end{figure}

\begin{figure}
    \centering
\includegraphics
[width=\textwidth]{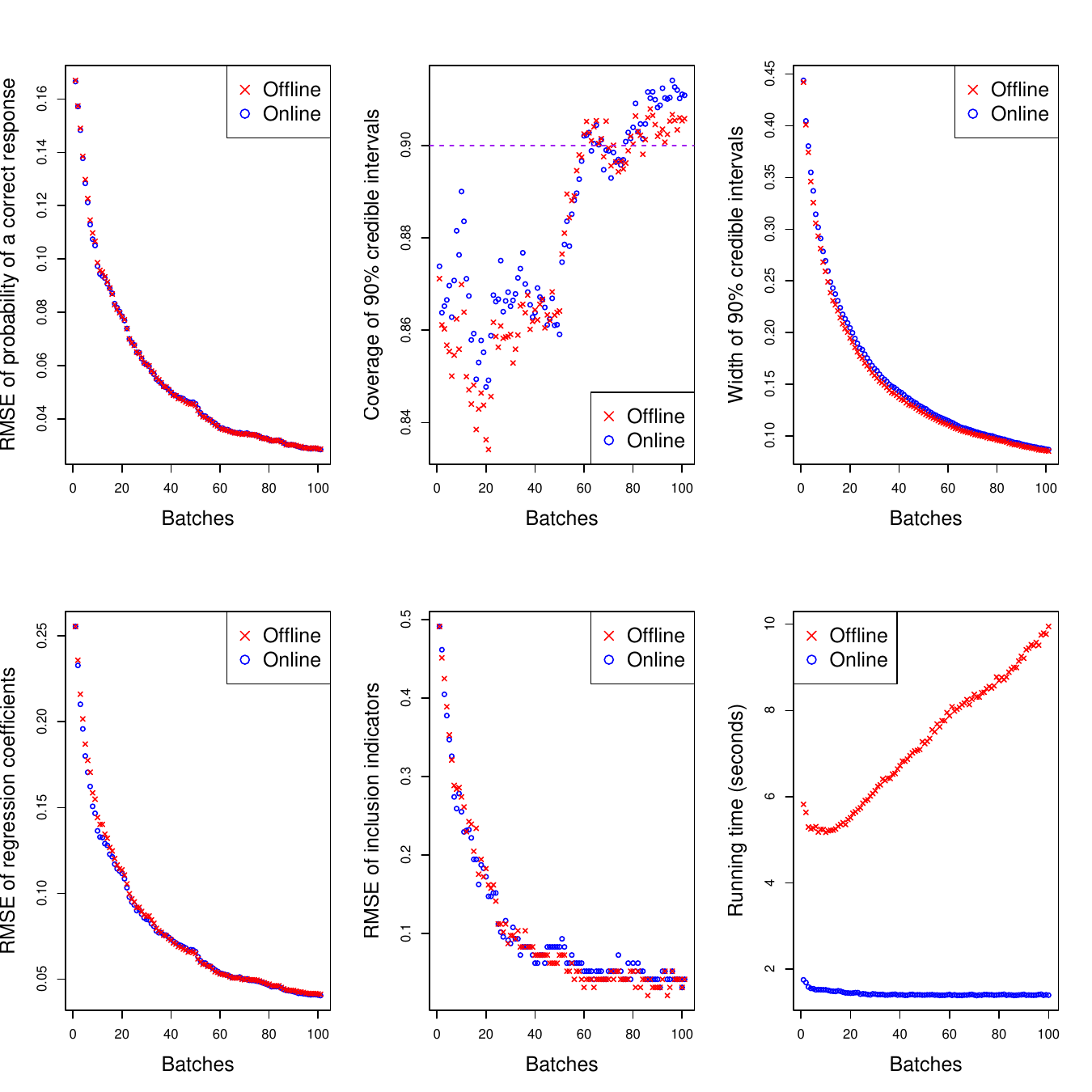}
    \caption{Simulation results for the independent nonsparse example, averaged over 25 replicates.}
    \label{fig:uncornonsparse}
\end{figure}

For prediction, we generate a new dataset with the same sample size as the training data, that is the test data sample size $N_{\rm test}=N=2000$. We estimate the posterior mean and the 5th and 95th percentiles of the distribution of predictive probabilities using Monte Carlo estimates, as shown in (\ref{eq:predictiveprob}), in Section \ref{subsec:statinf}. We compute i) the root mean squared error (RMSE) of the estimated posterior mean, ii) the empirical coverage of the resulting 90\% posterior credible interval formed with the quantiles, along with iii) the width of the 90\% credible interval, for evaluating the performance of the proposed method in the estimation of predictive probabilities and their associated uncertainty quantification. These three metrics are plotted in the top three panels of Figures \ref{fig:corsparse} to \ref{fig:uncornonsparse}. 

Suppose the vector of the true regression coefficients and the model indicator that generated the data are denoted by $\bbeta_{\mathrm{true}}$ and $\bm{\gamma}_{\mathrm{true}}$, respectively. 
iv) We estimate $\bbeta_{\mathrm{true}}$ by its Bayesian model averaged estimate given in (\ref{eq:bma-beta-mean}), and evaluate its accuracy with its RMSE. We estimate $\bm{\gamma}_{\mathrm{true}}$ using the median probability model (MPM) of \cite{Barb:Berg:2004} by thresholding the marginal posterior inclusion probabilities given in (\ref{eq:pip}) at 0.5. If a predictor's marginal posterior inclusion probability is 0.5 or more, it is included in the MPM; otherwise, the predictor is excluded from the MPM. v) We evaluate the accuracy of the MPM through RMSE. vi) Finally, we compare the running times 
of the offline and online BMA methods to evaluate the gain in speed. The metrics in (iv) to (vi) 
are plotted in the bottom three panels of Figures \ref{fig:corsparse} to \ref{fig:uncornonsparse}.

Figures \ref{fig:corsparse} to \ref{fig:uncornonsparse} compare the offline and online BMA methods in terms of prediction accuracy, estimation accuracy, uncertainty quantification, variable selection, and computational cost, under the four different simulation scenarios described earlier, with results averaged over 25 replicates. Overall, the online method closely resembles the offline method, while requiring substantially less time. The RMSE tends to decrease with increase in sample size, which is expected. The coverage varies from approximately 82\% to 92\% across the four simulation scenarios, and is generally closer to 90\% for larger sample sizes. It is well known that Bayesian 90\% credible intervals are not guaranteed to have exact frequentist coverage of 90\%, but the empirical coverage being generally close to 90\% for large sample sizes is promising. Predictably, we find that the width of the 90\% credible interval decreases with sample size. Additionally, we find the area under the ROC curve (not plotted here) varies  approximately from 0.68 to 0.89, on average, across the four different scenarios and 101 batches, suggesting that the methods have reasonably good predictive power for out of sample prediction.
Finally, the running time for online BMA stays fairly stable for each new batch update. However, it keeps steadily increasing approximately linearly for offline BMA as it reanalyzes the entire cumulative data after each batch arrival.

To summarize, the results from the simulation study demonstrate that we have been successful in designing an online BMA method that reduces the computational cost compared to offline BMA, without much loss in accuracy.

\clearpage
\normalsize
\section{Junyi Learning Analytics}\label{sec:application}

We apply the proposed online BMA method to data from the Junyi Academy \citep{JunyiOnlineLearningDataset}, an online platform that provides mathematics resources for K--12 students. The data contains student and problem
characteristics and their interactions, with a binary response indicating whether a problem was answered correctly. See \citet[Table~6]{nguy:dissertation:2026} for the codebook. Our online BMA approach is well suited to this sequential setting, where multiple predictors may contain related information and both model and parameter uncertainty are of interest. Our goals are to (1) identify predictors associated with  whether a student's answer is correct or not, (2) predict the success probability that is the probability that a student answers a new question correctly, (3) quantify the uncertainty of the predictive probability in (2) with an interval estimate, and  (4) assess how closely online BMA reproduces offline BMA inference at a lower computational
cost.

The predictor set consists of 18 variables, capturing student and problem characteristics, prior performance, and learning activity. Examples include problem  difficulty level and historical correct-response rates; student grade level, login activities, badges earned, accumulated energy points, and correct response rates. Several predictors, such as  correct response rate and recent activity measures, summarize a student's performance and engagement up to the time of each response.

\begin{figure}[h!]
    \centering
\includegraphics[width=.6\textwidth]{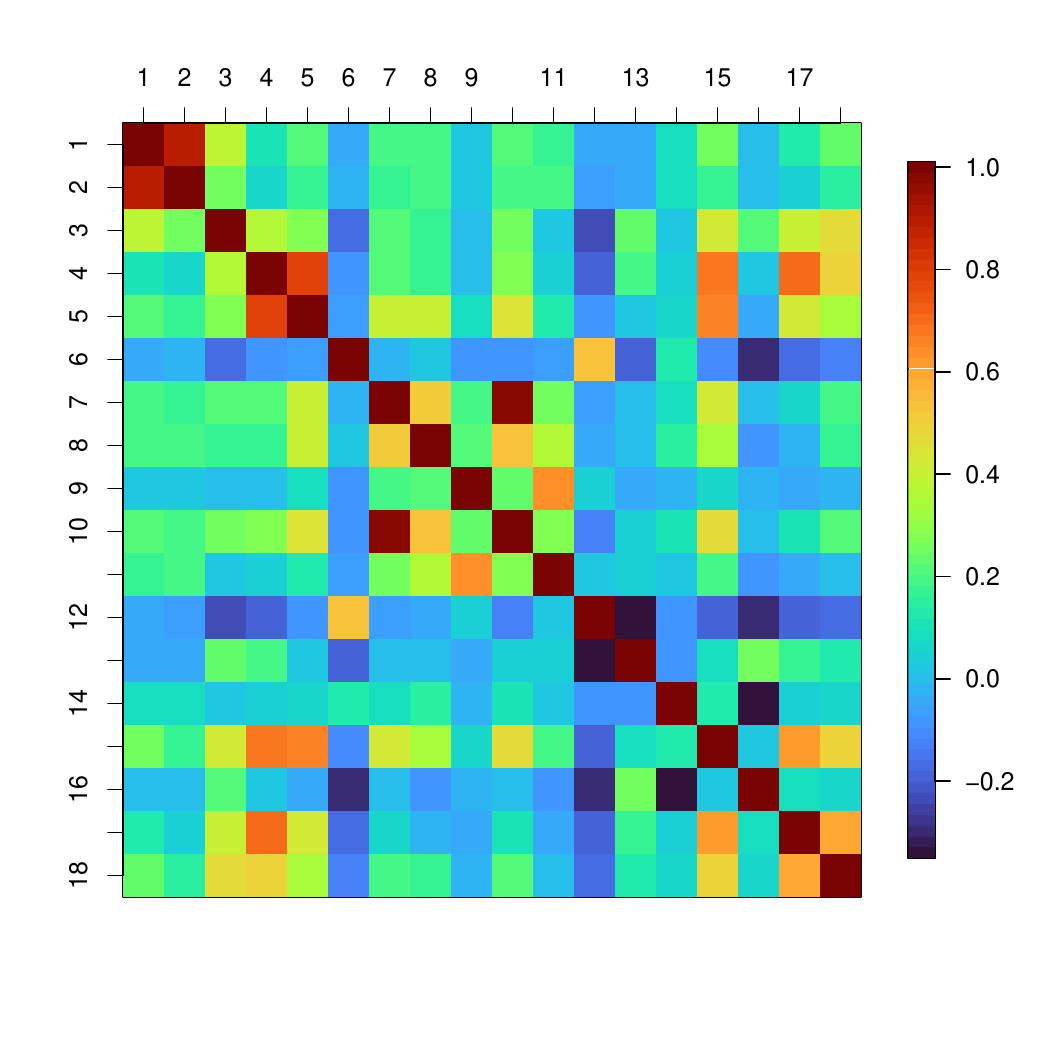}
\caption{Correlation structure of the predictors for Junyi dataset after the final batch. }
\label{fig:junyi-corrX}
\end{figure}

 Figure~\ref{fig:junyi-corrX} shows the predictor correlation matrix after 1501 batches, that is the entire training data. Examination of correlation matrices at previous batches shows a largely stable structure over time.  Several moderate to strong correlations are evident. For example, \texttt{level\_prev} ($X_3$) is moderately correlated with multiple predictors, including the highly-correlated pairs \texttt{problem\_number} and \texttt{exercise\_problem\_repeat\_session} ($X_1$ and $X_2$), and \texttt{points\_prev} and \texttt{badges\_cnt\_prev} 
($X_4$ and $X_5$). These dependencies reflect overlapping information about students' learning histories and create competing predictor configurations that motivate accounting for model uncertainty when estimating the student's probability of answering the current question correctly.

\begin{table}[ht]
\centering
\caption{Posterior summaries for the four highest probability models under
offline and online BMA. Coefficient entries are posterior medians with
90\% credible intervals in parentheses, conditional on a specific
model. Predictors excluded from a model have coefficient zero, which are omitted and shown with a dash.}
\label{tab:bma-both-ci}
\footnotesize
\setlength{\tabcolsep}{3pt}

\begin{tabular}{llcccc}
\toprule
& & Model 1 & Model 2 & Model 3 & Model 4\\
\midrule

Posterior prob.
& Offline & 0.24 & 0.21 & 0.11 & 0.04 \\
& Online  & 0.20 & 0.18 & 0.10 &  0.04 \\
\midrule

Intercept
& Offline
& $-3.83\,(-4.36,-3.33)$
& $-4.04\,(-4.52,-3.44)$
& $-3.64\,(-4.24,-2.99)$
& $-3.61\,(-4.35,-3.10)$
\\
& Online
& $-3.76\,(-4.39,-3.27)$
& $-4.05\,(-4.57,-3.39)$
& $-3.64\,(-4.26,-2.99)$
& $-3.51\,(-4.18,-3.04)$
\\[4pt]

\texttt{level\_prev}
& Offline
& $0.17\,(0.06,0.28)$ & -- & --
& $0.13\,(0.05,0.24)$ 
\\
& Online
& $0.15\,(0.05,0.25)$ & -- & --
& $0.16\,(0.05,0.26)$ 
\\[4pt]

\texttt{learning\_}
& Offline
& -- & -- & $-0.42\,(-0.72,-0.19)$
& $-0.40\,(-0.59,-0.13)$ 
\\
\texttt{stage\_junior}& Online
& -- & -- & $-0.42\,(-0.73,-0.13)$
& $-0.39\,(-0.62,-0.14)$ 
\\[4pt]

\texttt{correct\_rate}
& Offline
& $2.60\,(2.07,3.21)$
& $2.81\,(2.25,3.43)$
& $2.62\,(1.88,3.30)$
& $2.39\,(1.96,3.29)$
\\
& Online
& $2.57\,(1.96,3.28)$
& $2.89\,(2.26,3.53)$
& $2.62\,(1.98,3.29)$
& $2.43\,(1.66,3.11)$
\\[4pt]

\texttt{problem\_}
& Offline
& $4.84\,(4.26,5.47)$
& $5.02\,(4.41,5.49)$
& $4.78\,(4.26,5.36)$
& $4.80\,(4.32,5.35)$
\\
\texttt{correct\_rate}
& Online
& $4.80\,(4.21,5.36)$
& $4.98\,(4.45,5.55)$
& $4.78\,(4.24,5.31)$
& $4.76\,(4.34,5.34)$
\\

\bottomrule
\end{tabular}
\end{table}

\noindent \textbf{Streaming design.} 
We considered $2800$ observations, of which we reserve the last $1000$ for out-of-sample evaluation. This sample size allows repeated offline BMA analyses, in which cumulative data are reanalyzed after each batch, providing a benchmark for assessing how closely the proposed online BMA reproduces offline inference while reducing computing cost. We use the first $n_1=300$ observations as the initial batch and $n_b=1$ for  $b=2,\ldots, 1501$. At each update, the online BMA method uses only the current observation and summary information retained from previous batches.

One benefit of this small-batch design is that the posterior predictive distribution of $p(y_{b+1}=1 \mid \mathcal{D}_b, \bx_{b+1})$ is updated immediately after each new response, where $\bx_{b+1}$ contains the student's most recent learning and activity history and characteristics of a candidate problem. 
For example, the posterior mean of $p(y_{b+1}=1 \mid \mathcal{D}_b, \bx_{b+1})$ can  help predict and rank probabilities for a collection of candidate problems and identify those within a desired difficulty range. Importantly, the proposed BMA approach also provides credible intervals for these probabilities, which can  distinguish candidate problems with similar predicted success probabilities but different levels of uncertainty, and help to select problems whose predictive probability intervals lie within a desired range. 

 In the initial batches, two predictors were exactly collinear when the
accumulated sample size was still small. We therefore used a log~F$(4,4)$
prior for one of the affected coefficients, following \cite{Gree:Mans:2015} and \citet{ghosh:tan:luo:2025}. The prior admits a convenient data-augmentation representation: for the affected coefficient $\beta_j$, four pseudo-observations are appended, with two successes and two failures. For each such observation, the covariate vector is zero except for a 1 in the position
corresponding to $\beta_j$. 
Adding the 4 rows corresponding to the prior helps in removing the multicollinearity. Thanks to the form of the $\log F$ prior, maximizing the posterior is equivalent to maximizing the likelihood of the augmented data, allowing the renewable estimation algorithm to directly provide the quantities required for online BMA. The exact collinearity disappeared in later batches as more observations accumulated, but we retained the prior to maintain a common prior specification across batches, and its influence diminishes as data size increases.\\\vspace{.5cm}

\begin{figure}

\centering

\begin{minipage}[t]{0.37\textwidth}
\vspace{20pt}
\centering

    \begin{subfigure}[t]{\textwidth}
        \centering
        \includegraphics[width=\textwidth]
        {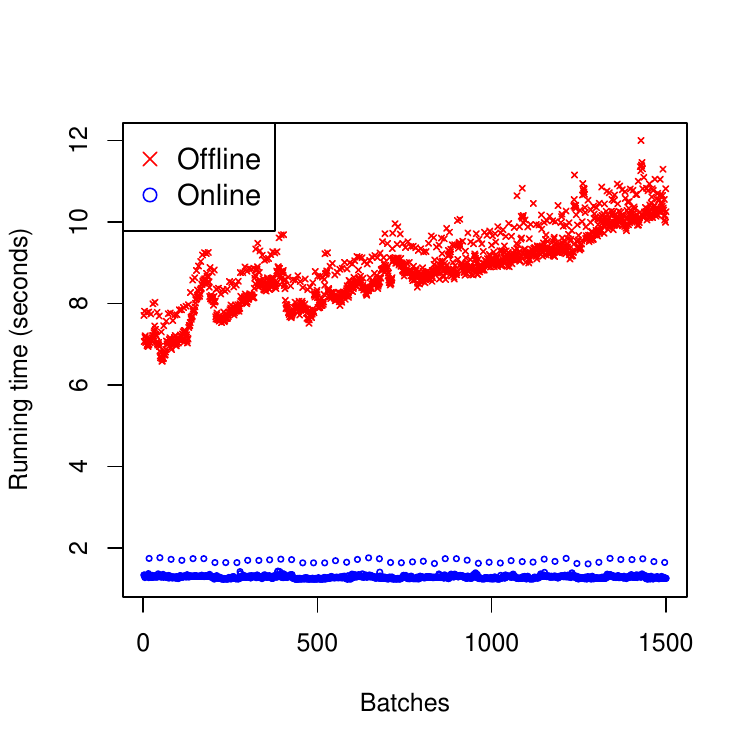}
        \setcounter{subfigure}{0}
        \caption{Running time per batch.}
        \label{fig:junyi-time}
    \end{subfigure}

    \vspace{.5em}

    \begin{subfigure}[t]{\textwidth}
        \centering
        \includegraphics[width=\textwidth]
        {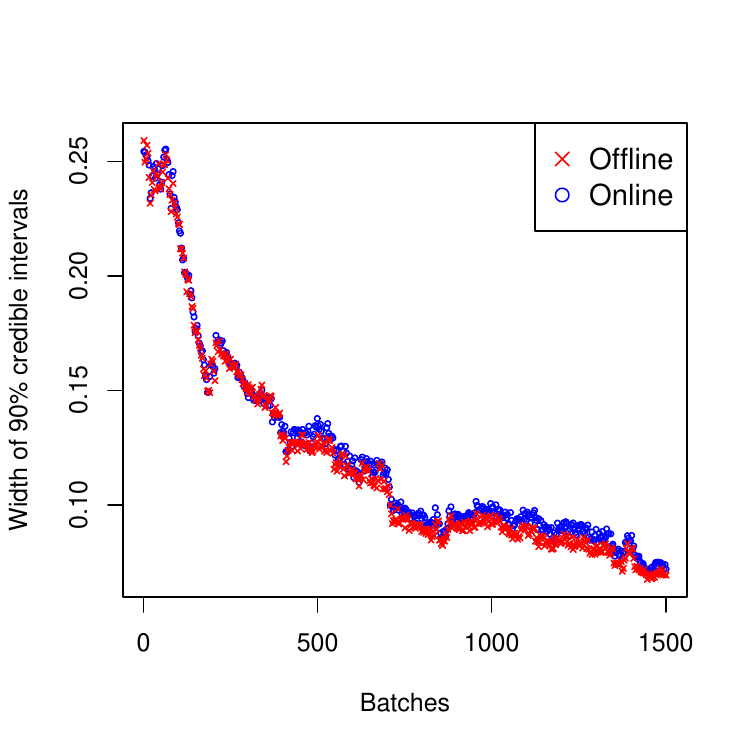}
        \setcounter{subfigure}{2}
        \caption{Width of 90\% CI for probabilities, averaged over test points.}
        \label{fig:junyi-ciwidth}
    \end{subfigure}

\end{minipage}
\begin{minipage}[t]{0.57\textwidth}
\vspace{20pt}
\centering

    \begin{subfigure}[t]{\textwidth}
        \centering
        \includegraphics[height=5.55in,width=\textwidth]
        {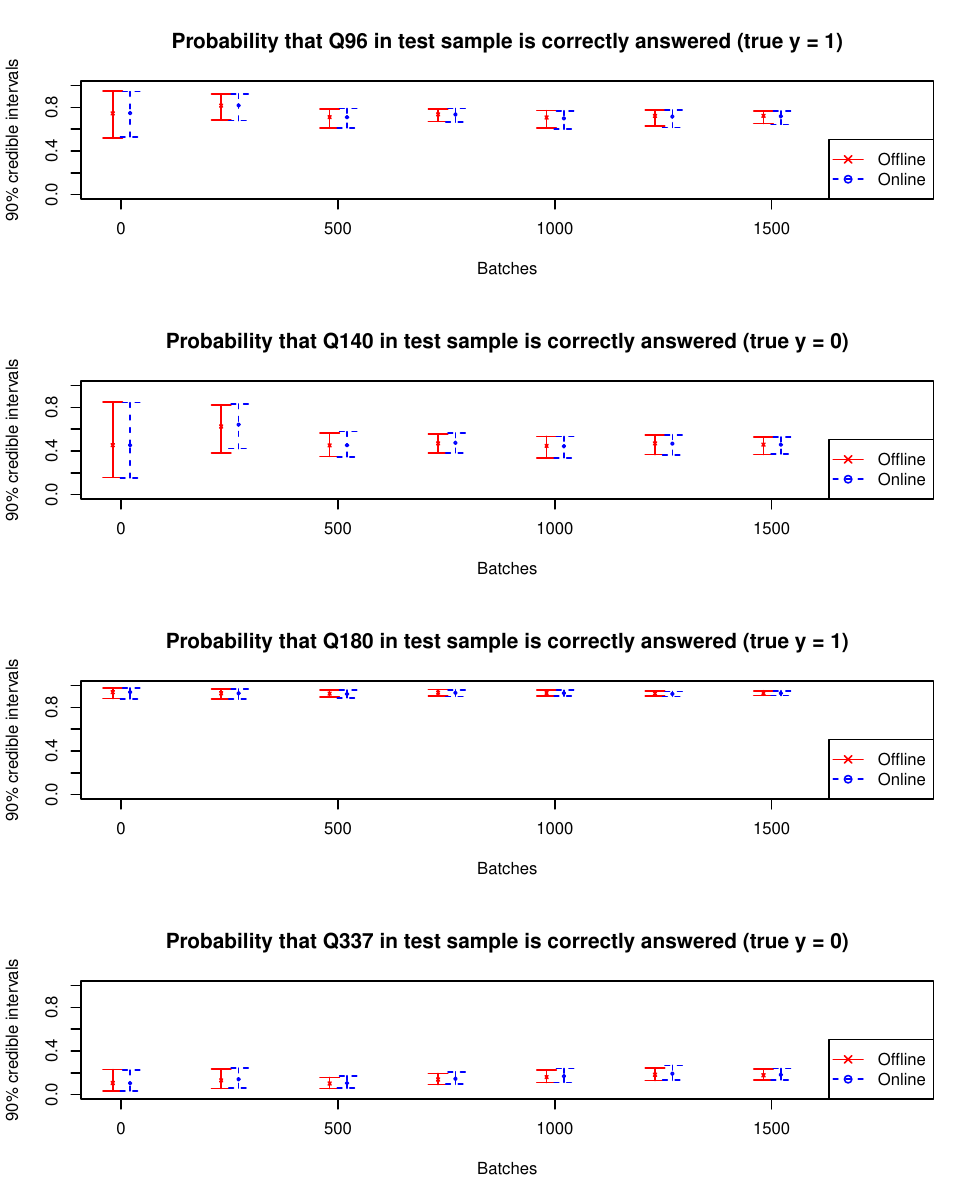}
        \setcounter{subfigure}{1}
        \caption{Point estimate and 90\% CI of correct rate, $p(\bx)$, for four representative questions.}
        \label{fig:junyi-probint}
    \end{subfigure}

\end{minipage}
\caption{
Comparing online and offline BMA over time. The periodic spikes in panel (a) are due to R’s automatic garbage collection, to free up memory used by temporary R objects.} \label{fig:junyi-3plots}

\end{figure}

\noindent \textbf{Analysis results.} 
We now assess how accurately online BMA approximates offline BMA and at what computing cost, and then illustrate the inference provided by BMA. 
In Figure~\ref{fig:junyi-3plots}, Plot~(a)  
shows that the computing time of online BMA per update remains low, and that of offline BMA
increases as more observations must be reanalyzed.
Despite this computational difference, the two methods produce remarkably
similar inference. Plot~(b) shows close agreement in posterior means and intervals for the predicted correct-response probabilities (shown for four representative problems).  More broadly, plot~(c) shows the widths of these intervals 
(averaged over all $1000$ test samples) closely track each other across batches. Predictive accuracy is also nearly identical, as AUC on the test set of the online BMA closely tracks that of the offline over time and both reach \(0.78\) at the final update (plot not shown).  In terms of model uncertainty and variable importance, the two approaches are also similar
at each time point. Shown in
Figure~\ref{fig:junyi-topmodels} is the close agreement in the
posterior probabilities of the leading models and the marginal posterior
inclusion probabilities (PIPs) at the final update. Finally, for regression coefficient inference, Table~\ref{tab:bma-both-ci} compares posterior medians and $90\%$ credible intervals conditional on each of the five highest-probability models, again showing close agreement between the two approaches. Overall, the analysis on Junyi data shows that online BMA closely reproduces offline in all aspects including variable selection, prediction, estimation, and the corresponding uncertainty quantification, while substantially reducing computing cost.

Given that the results from the two approaches closely agree, we only quote numbers from the online BMA to demonstrate BMA inference below. At the final update, no single model dominates the posterior:
the three most probable models have posterior probabilities $0.20$, $0.18$, and $0.10$, respectively, with the remaining $0.52$ distributed over a wide range of predictor configurations, illustrating the importance of accounting for model uncertainty rather than solely conditioning inference on a single selected model. 

Marginalizing over the model space provides one way to make inference on individual
predictors. The PIP summarizes the
evidence for including a predictor (Figure~\ref{fig:junyi-topmodels}). 
Variables \texttt{correct\_rate} and
\texttt{problem\_correct\_rate}, representing the student’s cumulative performance and the historical success rate of the problem, respectively, are the most strongly supported predictors,
both with PIPs essentially equal to one. The variable importance for
\texttt{level\_prev} and \texttt{learning\_stagejunior} are moderate,
with PIPs of $0.45$ and $0.30$,  
respectively. 
As discussed in the methodology section, interpretable  assessment of predictor effect size and their uncertainty can be obtained conditional on a highly probable model, typically through a conditional posterior mean or median and credible interval. As shown in Table~\ref{tab:bma-both-ci}, conditioning on the highest probability model, the posterior median coefficients (90\% credible interval) of \texttt{correct\_rate}, \texttt{problem\_correct\_rate} and \texttt{level\_prev} are
 $2.57 (1.96, 3.28)$, $4.80 (4.21, 5.36)$ and $0.15 (0.05, 0.25)$, respectively. This for example implies that a $27$-percentage-point
increase in the student's cumulative \texttt{correct\_rate} is associated with doubling
the odds of a correct response, since
$\exp(2.57\times0.27)\approx 2.00$, or equivalently, $\log(2)/2.57 \approx 0.27$; and a similar effect on the outcome can be achieved by offering an easier problem that has a $15$-percentage-point higher \texttt{problem\_correct\_rate}, since
$\exp(4.80\times0.15)\approx 2.05$. The credible intervals conveniently quantify uncertainty in these effect sizes. In particular, dividing $\log(2)$ by the endpoints of the coefficient credible intervals gives a 90\% credible range for the increase associated with a doubling of the odds: approximately 21--35 percentage points for \texttt{correct\_rate} and 13--16 percentage points for \texttt{problem\_correct\_rate}. 
 In addition, the third and last predictor in the top model is \texttt{level\_prev}  (0--4), representing progressive student proficiency. It has a conditional posterior median of $0.15$ and a $90\%$ credible interval of $(0.05,0.25)$. Thus, conditional on the top model, a one-unit increase in \texttt{level\_prev} is associated with a 16\% increase in the odds of a correct response, with a 90\% credible interval for the odds ratio of $(\exp(0.05),\exp(0.25))\approx(1.05,1.28)$.

\clearpage

\section{Summary}
In this article, we proposed an online BMA methodology that allows joint 
uncertainty quantification for the choice of included
variables in the model, along with the uncertainty quantification of regression coefficients within 
a model, such that inference can be updated at each time point of the data stream. Specifically, the proposed work introduced an online Bayesian framework for logistic regression that simultaneously achieves three goals that have not been jointly addressed in existing methods: computationally efficient analysis of streaming data, uncertainty quantification for predictive probabilities using interval estimates, and interpretable
assessment of predictor importance and regression effects through PIP and model-specific coefficient inference. The proposed methodology advances the state of the art in streaming data analysis and has broad relevance for psychometrics, educational measurement, and related domains. 

We evaluated the proposed methodology through both simulation studies and an empirical application in learning analytics. In the simulation studies, we examined scenarios with independent and correlated predictors, varying signal strengths and sparsity patterns. Performance between online and offline BMA were assessed in terms of predictive accuracy, coverage of credible intervals for predictive probabilities,   
variable selection, regression coefficient estimation and computational efficiency. Overall, the proposed online BMA method produced results very similar to those of its offline counterpart, in both the simulation studies and the learning analytics application, while needing only a fraction of the computing cost. The learning analytics dataset and the associated R code used for its analysis are provided at \url{https://github.com/joyeeghosh/online-BMA-joint-inference}.

\paragraph{Use of Generative Artificial Intelligence (AI)}
ChatGPT was used to check code, produce code for Figure \ref{fig:junyi-topmodels}, and for language editing. All content
was subsequently reviewed and edited by the authors, who assume responsibility for the final manuscript.

\bibliography{ref,jcgs}

\end{document}